\documentclass[onecolumn,authoryear]{els-mrw} 

\usepackage{amsmath,amssymb,amsfonts,amsthm,makeidx,graphicx}
\usepackage{txfonts}
\usepackage{helvet}
\newcommand{\actaa}{Acta Astron.}		
\newcommand{\aj}{Astron. J.}                   
\newcommand{\araa}{Annu. Rev. Astron. Astrophys.} 
\newcommand{\apj}{Astrophys. J.}               
\newcommand{\apjl}{Astrophys. J. Lett.}        
\newcommand{\apjs}{Astrophys. J. Suppl. Ser.}  
\newcommand{\aap}{Astron. Astrophys.}          
\newcommand{\aaps}{Astron. Astrophys. Suppl. Ser.} 
\newcommand{\mnras}{Mon. Not. R. Astron. Soc.} 
\newcommand{\pasp}{Publ. Astron. Soc. Pac.}    
\newcommand{\pasa}{Publ. Astron. Soc. Australia}   
\newcommand{\pasj}{Publ. Astron. Soc. Japan}   
\newcommand{\nat}{Nature}                      
\newcommand{\na}{New Astron.}                  

\begin{document}

\chapter{The Milky Way Bulge}\label{chap1}

\author[1,2]{Manuela Zoccali}%
\author[3,4]{Elena Valenti}%

\address[1]{\orgname{Pontificia Universidad Cat\'olica de Chile}, \orgdiv{Instituto de Astrof\'isica}, \orgaddress{Av. Vicu\~na Mackenna 4860, 782-0436 Macul, Santiago, Chile}}
\address[2]{\orgname{Millennium Institute of Astrophysics},  \orgaddress{Av. Vicu\~na Mackenna 4860, 782-0436 Macul, Santiago, Chile}}
\address[3]{\orgname{European Southern Observatory}, 
\orgaddress{Karl Schwarzschild-Stra{\ss}e 2, D-85748 Garching bei M{\"u}nchen, Germany}}
\address[4]{\orgname{Excellence Cluster ORIGINS}, 
\orgaddress{Boltzmann-Stra{\ss}e 2, D-85748 Garching bei M{\"u}nchen, Germany}}

\articletag{Chapter Article tagline: update of previous edition,, reprint..}

\maketitle

\begin{glossary}[Glossary]
\term{[Fe/H]}: The stellar metallicity, defined as the ratio between the number density of Fe over H atoms, measured in a given star, with respect to the solar value: i.e., [Fe/H] = log(N$_{Fe}$/N$_{H}$) $-$ log(N$_{Fe}$/N$_{H}$)$_{\odot}$. \\
\term{APOGEE}: Apache Point Observatory Galactic Evolution Experiment is a high\--resolution (R$\sim$22,000) infrared spectroscopic survey carried out in the Northern (APOGEE\--N) and Southern (APOGEE\--S) Hemisphere \citep{majewski+2017,APOGEE-N}  \\
\term{ARGOS}: Abundances and Radial velocity Galactic Origins Survey is a medium\--resolution (R=11,000) optical spectroscopic survey of giants in the Milky Way bulge and disk \citep{ARGOSsurvey}\\ 
\term{DENIS}: DEep Near Infrared Survey provides I, J and K band photometry of the Southern sky \citep{DENISsurvey}\\
\term{DIRBE}: Diffuse Infrared Background Experiment is a 10\---band absolute photometer, onboard NASA's Cosmic Background Explorer (COBE) satellite, covering the wavelengths between 1 and 300 microns \citep{DIRBE}\\
\term{Gaia ESO}: The Gaia ESO Survey is a medium\--resolution (R$\geq$10,000) optical spectroscopic survey of K giants in the Milky Way bulge, disc and clusters \citep{GESsurvey} \\
\term{GIBS}: Giraffe Inner Bulge Survey is a medium\--resolution spectroscopic survey in the optical band targeting K giants in the Milky Way bulge \citep{GIBSsurvey}\\
\term{OGLE}: Optical Gravitational Lensing Experiment is a multi\--epochs optical survey of the Milky Way bulge and Magellanic Clouds (https://ogle.astrouw.edu.pl/)\\
\term{PIGS}: Pristine Inner Galaxy Survey is a metallicity-sensitive narrow-band CaHK photometric survey targeting metal\--poor stars in the Milky Way inner region \citep{PIGSurvey}\\
\term{VVV}: VISTA Variables in the V\'\i a L\'actea Survey is a near infrared multi\--epochs photometric survey of the Mikly Way bulge and disk \citep{VVVsurvey}\\
\end{glossary}

\begin{glossary}[Nomenclature]
\begin{tabular}{@{}lp{34pc}@{}}
BW & Baade Window \\
CMD & Color-Magnitude Diagram \\
GC & Globular cluster \\
IMF & Initial Mass Function \\
m-poor & metal-poor \\
m-rich & metal-rich \\
MDF & Metallicity Distribution Function \\
MW & Milky Way \\
MS\--TO & Main Sequence Turn-Off \\
PopI/PopII & Population I / Population II \\
RC & Red Clump \\
\end{tabular}
\end{glossary}

 \begin{abstract}[Abstract]
This chapter reviews the three-dimensional structure, age, kinematics, and chemistry of the Milky Way (MW) region within $\sim$2 kpc 
from its center (hereafter referred to as the 'bulge') from an observational perspective. While not exhaustive in citations, this 
review provides historical context and discusses the main controversies and limitations in the current consensus. The nuclear bulge 
region, within  $\sim$200 pc from the Galactic center, has been excluded from this review. This very complex region, 
hosting dense molecular clouds and active star formation, would deserve a dedicated paper.
 \end{abstract}

\begin{glossary}[Key Points]
\begin{itemize}
\item 
  The central region of the MW Galaxy, within $\sim$2 kpc from the center, here called "the bulge", is a massive stellar component 
  including 1.3-2$\times$10$^{10}$M$_\odot$. It makes up about one quarter of the total stellar mass of the Galaxy.
\item 
  The Galactic bulge includes a metal poor (m-poor) and a metal rich (m-rich) component. They differ in several key properties.
\item  
  The m-poor component is made up exclusively by old stars ($\sim$10 Gyr), with mean [Fe/H]$\sim$$-$0.5 dex and mean 
  [$\alpha$/Fe]$\sim$+0.25. It is largely dominant far from the midplane (Z$\gtrsim$0.8), and it has a roughly spheroidal shape.
  Its origin is not firmly established.
\item
  The m-rich component is also made up by old stars, but it might include some poorly constrained fraction (0-35$\%$) of
  intermediate age stars (2-8 Gyr). It has mean [Fe/H]$\sim$+0.4 dex and mean [$\alpha$/Fe]$\sim$0, it is dominant near 
  the midplane and it has a bar shape, with a Boxy/Peanut in its outer region. As every bar, it is the product of dynamical
  instabilities of the Galactic disk, induced by spiral arms.
\end{itemize}
\end{glossary}


\section{Introduction}\label{intro}

\begin{figure}[h]
\centering
\includegraphics[width=0.7\textwidth]{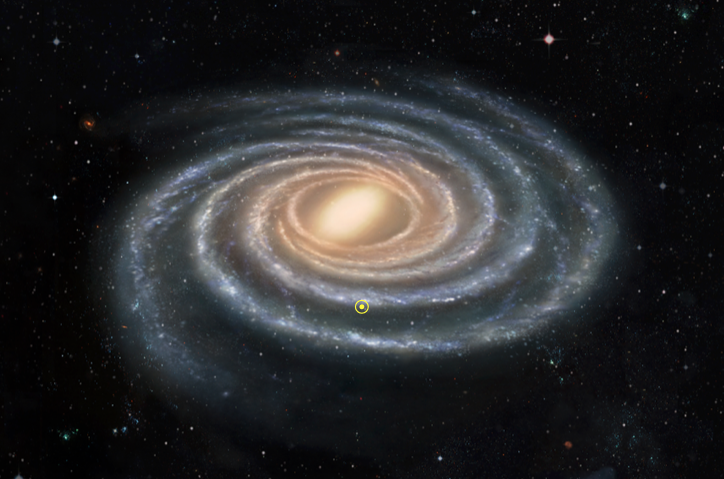}
\caption{Artist view of the MW galaxy, as seen from an oblique angle. The location of the Sun, within the disk, is
shown as a yellow circled dot. Credit: Xing-Wu Zheng \& Mark Reid  BeSSeL/NJU/CFA.}
\label{fig:mw_art}
\end{figure}

With a total mass of M$\sim$10$^{12}$ M$_\odot$ \citep{BlandHawthornGerhard16}, the MW galaxy is a giant spiral with four spiral arms 
extending from the core to the outer region, hosting a prominent bar at the center (Fig.~\ref{fig:mw_art}). Its morphological type is 
close to SBb/SBbc, although its exact shape has been, and still is, very hard to derive from our observing position. From the Earth, 
located in the middle of the disk, it is  particularly difficult to explore the Galaxy's central region, as the line of sight needs 
to cross more than 8 kpc of dust and stars. The dust attenuates and reddens the flux of sources behind it, while foreground disk stars 
need to be recognized as such, in order to be removed from the sample of bulge stars under study. In addition, the innermost few degrees 
of our Galaxy host a huge number of stars, making confusion a very serious problem. Despite these challenges, the MW region within 
$\sim$2 kpc from the center, that we will call here the "bulge", is the only spiral bulge that can be resolved in stars down to the 
lower main sequence. For the closest external bulge, that of Andromeda, we can barely resolve Main Sequence Turn Off (MS-TO) stars, 
and only in its outer region, while high resolution spectroscopy is currently still out of reach. The MW bulge includes between 1/3 
and 1/4 of the Galaxy's total stellar mass (see Section~\ref{sec:mass}), mostly in the form of $\sim$10 Gyr old stars (see Sec.~\ref{sec:age}). 
Most of the remaining MW mass is in the disk, on average younger than the bulge, while the halo, coeval or even older than the bulge, 
makes up only $\sim$1$\%$ of the MW total stellar mass. Therefore, the MW bulge is the first massive component to form stars, and the 
only galaxy bulge that we can explore in detail, hence its relevance in the context of galaxy formation.

\section{The three-dimensional structure of the Galactic bulge}\label{sec:3D}

\subsection{Early evidences for a central bar}   

The presence of a large concentration of stars towards the Sagittarius constellation is very evident to the naked eye, in a dark night. 
After \citet{shapley30} analyzed the space distribution of GCs, concluding that the center of the Galaxy was in that direction, it became 
generally accepted that the so-called Sagittarius great star clouds were located near the Galactic center, forming a bulge qualitatively 
similar to that of Andromeda.

The first large scale map of the MW spiral structure was constructed by \citet{Oort59} by means of HI 21-cm observations at Leiden and 
Sydney. By comparing the spiral pattern with that of external galaxies, \citet{deVaucouleurs64} proposed that our Galaxy was a SAB type, 
at the transition between a normal and a barred galaxy. Although the region inside $\sim$5 kpc was completely missing from the map, he 
argued that the high multiplicity of the MW spiral arms was more consistent with that of spiral galaxies with a weak bar, such as 
NGC~6744 or NGC~4303. Note that the existence of a correlation between the spiral shape and the Hubble type and/or the presence of a bar 
is currently debated \citep[e.g.,][and references therein]{smith+24}.

\begin{figure}[t]
\centering
\includegraphics[width=\textwidth]{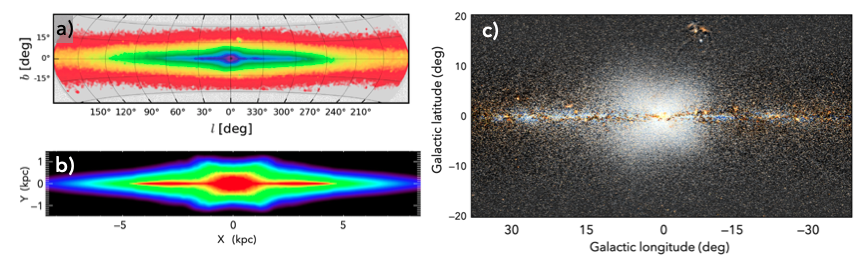}
\caption{The MW bar and B/P. Panel a): unWISE map of the MW, showing the well known asymmetry in the central region (blue contours), 
due to the presence of the bar \citep{mosenkov+21}. Panel b): model of the B/P and long bar, fitted to the data, by \citet{wegg+15}. 
Panel c): WISE map, with the median of each row of the image subtracted, in order to provide a better 
contrast and highlight the B/P, or X-shape, structure. Credit: NASA/JPL-Caltech/D. Lang. }
\label{fig:3D}
\end{figure}

Improved HI observations, now including the inner Galaxy, carried out in 1978 from Green Bank, allowed \citet{liszt+80} to model the 
motion of the gas within 2 kpc from the Galactic center, and conclude that it was consistent with a bar-like structure with its near 
end at positive longitudes. This result was later confirmed by \citet{binney+91}, \citet{englmaier+99} and \citet{fux+99}.
The presence of a bar was independently confirmed also by \citet{blitz+1991} and \citet{weiland+94}, who analyzed the integrated near-IR 
flux from the balloon observations by \citet{matsumoto+82} and from the COBE DIRBE maps, respectively. Both of them found that the flux contours were 
larger at positive than at negative longitudes, qualitatively consistent with the presence of a triaxial bar with its near end in the first 
Galactic quadrant (0$^\circ$$<$$\ell$$<$90$^\circ$). Special mention is deserved by the work of \citet{dwek+1995}, also based on COBE DIRBE 
data. Beside confirming the barred structure, with both axis ratio and orientation exceptionally accurate, compared to modern measurements, 
they also derived the first photometric mass of the bulge/bar (M$_{\rm B}$=1.3$\times$10$^{10}$ M$_\odot$) from the total L$_{\rm K}$ 
luminosity (L$_{\rm K}$=4.1$\times$10$^8$ L$_\odot$). The left-right asymmetry in the integrated IR flux of the central MW is very clearly 
visible in the more recent allsky survey unWISE \citep{meisner+17} shown in Fig.~\ref{fig:3D}a, and analyzed by \citet{mosenkov+21}.

\subsection{The bar as traced by Red Clump stars}   

The final proof of the presence of the bar, together with a more robust characterization of its structural parameters, came with the analysis 
of bulge RC stars across different directions. The idea behind this method takes advantage of the very little sensitivity of the RC magnitude 
to age (see Sec.\,\ref{sec:age}). 
The dependence upon metallicity is slightly larger, but still mild within the metallicity range spanned by bulge stars, with the exact coefficient 
between RC absolute magnitude and [Fe/H] decreasing from optical to near IR bands \citep{wang21}. Therefore, the peak apparent magnitude of RC stars 
can be used as distance indicator, as it will be brighter in the direction where stars are, on average, closer to the Sun, and fainter where they 
are further away. This method is especially robust in mapping the MW bar, because possible population effects, such as metallicity or age gradients, 
are very unlikely to explain the observed asymmetry between positive and negative longitudes. Interstellar extinction, instead, poses a more serious 
challenge. In principle, the mean color of the RC gives information about the relative extinction (reddening) and a suitable extinction law allows 
one to convert it into an absolute extiction in the selected photometric band. In practice, however, the presence of longitude asymmetries in the 
mean extinction implies that the derived bar parameters are mildly affected by the choice of the extinction law.

Following \citet{stanek+94, stanek+97}, dozens of studies found that RC stars at positive longitudes are brighter than their counterpart at negative 
longitudes \citep{lopez-corredoira+97, unavane+98, bissantz+02, babusiaux+05, benjamin+05, nishiyama+05}.
While most of the early works analyzed the magnitude of RC stars across a few line of sights, at different longitude, with the advent of surveys it 
became possible to map RC stars across the whole bulge area, thus deriving the bar shape parameters via a thorough comparison with triaxial bar models. 
The first such wide area study was conducted by \citet{rattenbury+07}, who analyzed the observed magnitude of RC stars in 44 bulge fields, across 
$\sim$11 square degrees, from the OGLE-II optical survey. An update was later provided by \citet{cao+13}, based on OGLE-III data, across $\sim$90 
square degrees, within the range $-$10$^\circ$$<$$\ell$$<$10$^\circ$ and 2$^\circ$$<$b$<$7$^\circ$. More recent determinations were based on near IR 
data from the VVV survey, thus 
reaching the more extincted and denser region close to the Galactic midplane \citep{wegg+13, simion+2017, paterson+20}. The use of near IR bands also 
minimized the effect of extinction on the derived bar parameters. Incidentally, the possibility to map the color of RC stars across the whole bulge 
allowed the construction of high spatial resolution extinction maps towards this region, useful for a variety of other applications 
\citep[e.g.,][]{gonzalez+12, schultheis+14redd, surot+20redd}.

In principle, four main parameters describe the bar shape: the scale length of the semi-major axis; the ratio between the minor axis (projected onto 
the plane) and the major one; the ratio between the vertical minor axis and the major one, and the pivot angle between the major axis and the 
Sun--Galactic center direction. The semi-major axis scale length has been estimated to be $\sim$1 kpc, although this parameter is the most difficult 
one to measure, as it is prone to strong biases, depending on the method to constrain it \citep{ghosh+24}. The axis ratios are more robust, and 
although the precise values vary between different studies, typical ratios of the two semi-minor axes (b and c) relative to the major one (a) are 
(1:b/a:c/a)=(1:0.4:0.3). The pivot angle has been measured between 20 and 30 degrees, with a peak around $\sim$27 degrees. 

There is relatively good agreement, in the literature, concerning the shape parameters of the bar. Nonetheless, as recently discussed by 
\citet{hey+23} and \citet{zoccali+24RRL}, the errors on the distances can have a non-negligible impact on the bar shape. Errors are present only 
along the line of sight, because the coordinates have virtually zero uncertainty, hence they stretch the major axis of the bar more than the two 
minor ones, artificially aligning the bar with the $\ell$=0$^\circ$ direction. The magnitude of the effect depends, of course, on the precision of the 
distance indicators: for a relative error of $\sim$10$\%$, as in the case of RC stars, this has a non negligible impact, never modeled and 
corrected so far.

The MW bar has a Boxy/Peanut (B/P), often called X-shape, in its outer region. The feature that revealed its presence is a magnitude split 
of the RC, in fields at $\ell$=0$^\circ$ and $|b|$$>$5$^\circ$ \citep{mcwilliam+10, nataf+10, saito+11}. The B/P was confirmed and characterized
by several studies on independent sets of data \citep[e.g.,][]{wegg+13}, including the unWISE allsky map, showing this feature quite clearly 
\citep[Fig.~\ref{fig:3D}c;][]{ness+2016_x}. This structure, together with a thin long bar extending up to $\sim$5 kpc in the Galactic plane 
\citep[Fig.~\ref{fig:3D}b;][]{wegg+15} was predicted long ago by dynamical models of barred spirals \citep{pfenniger+91, patsis+02, athanassoula05}, 
as the results of either buckling \citep[e.g.,][and references therein]{collier20} or vertical resonance heating \citep{pfenniger+91}. Both 
phenomena affect only the central part of bars, converting the main elliptical orbits supporting the bar (called $x_1$) into "banana-shaped" 
orbits ($x_1 v_1$) supporting the Boxy/Peanut, but leaving the outer bar unchanged, hence its thin extension. Such complex shapes are found in 
about half of local edge-on spirals \citep{lutticke+00,laurikainen+14}.

\subsection{The role of pulsating variables}   
Pulsating variables, such as RR Lyrae (RRL), Miras and Cepheids, are widely used as distance indicators, because they follow a tight period-luminosity-metallicity 
relation. Therefore they have played an important role in the study of the bulge 3D shape. In addition of being among the most precise distance indicators, 
different types of pulsating variables are stars in specific age ranges, thus allowing us to dissect the bulge shape as a function of formation epoch. 

\subsubsection{RR Lyrae variables} \label{rrl}   
RRL are probably the most abundant ($>$30,000), and certainly the most widely studied variables in the Galactic bulge. They are core He burning stars 
crossing the instability strip during the horizontal branch phase. As such, they represent a clean tracer of the oldest population ($>$10 Gyr) and are 
also much more abundant in the m\--poor stellar component. Indeed, younger stars burn He confined in the RC, while old m\--rich stars may cross to 
the blue only if they loose a large fraction of their envelope. RRL are relatively easy to identify, as they are short period (0.3-1 days), and 
relatively bright stars (M$_V$$\sim$0.5 mag).
Public catalogs of bulge RR Lyrae have been provided by the OGLE optical survey \citep{ogleIV}, the Gaia DR3 \citep{clementini+22}, and the VVV near 
IR survey \citep{dekany+20, molnar+22, zoccali+24RRL}. Different catalogs obviously have different degrees of purity and completeness, as isolating 
bona fide variables of any kind, from a huge number of time series photometry, is all but a trivial task. Attempt at tracing the old bulge component 
by using one or more of the above sources gave conflicting results. Some authors found an axi-symmetric structure \citep[e.g.,][]{dekany+13}, while 
others found a shape compatible with that of the main Galactic bar \citep[e.g.,][]{pietrukowicz+15}. Other authors suggest that we might be seeing a 
mix of two populations, with the most m\--poor one more axi-symmetric than the m\--rich one \citep[e.g.][]{du+20}, while others suggest that the observed 
discrepancy might be due to the uneven sampling of RRL across the longitude axis, due to the asymmetric distribution of interstellar clouds \citep{zoccali+24}. 
A recent review by \citet{kunder22} discusses how 
the kinematics of RR Lyrae can help us establish whether they belong to a rotating structure such as a bar, or a pressure supported spheroid. Along 
the same lines, \citet{olivares+24} used six dimension phase space information for some of these variables, recently available from the combination 
of radial velocities and proper motions, to model their orbits and isolate bona fide bulge RR Lyrae from halo or thick disk interlopers. Unfortunately 
the problem is not yet solved, and it will most likely require all of the above: a robust classification of RR Lyrae, a proper treatment of extinction, 
and 3D kinematics plus orbit analysis.

\subsubsection{Type II Cepheids}    
These are also exclusively old and preferentially m\--poor stars, in an evolutionary phase posterior to core He burning. With respect to RRL they have 
longer periods (1–80 days), and are 1–3 mag brighter. They also follow a period-luminosity relation that makes them good distance indicators. Almost 
1000 Type II Cepheids have been identified and studied in the Galactic bulge \citep{braga+18, griv+21}. The resulting spatial distribution is 
ellipsoidal, although less elongated than the Galactic bar, as traced by RC stars.

\subsubsection{Mira variables}   
Miras are fundamental mode pulsating variables at the tip of the Asymptotic Giant Branch \citep[see][for a recent review]{huang24}. As such, they 
are very bright and ubiquitous, since all the stars with masses 0.8$_\odot$$<$M$<$8M$_\odot$ experience this phase. They come in two flavors: O-rich 
and C-rich stars, depending on the dominant element in their envelope. O-rich stars, with periods lower than 400 days, follow a tight period-luminosity 
relation in near IR, and they also follow a period-age relation \citep{trabucchi+22}, though with a non-negligible age spread at any given period. 
Thanks to the above, Mira variables, in principle, allow us to slice a given Galactic component into age bins. In practice, though, Miras are hard 
to study because they require long time baselines. Also, their light curves show periodic variations due to the pulsation, mixed with long time 
modulations due to periodic obscurations of their atmosphere by mass loss. Nonetheless, bulge Miras from Gaia DR2 have been analyzed by \citet{grady+20}, 
who found that the oldest ones (9-10 Gyr) trace an axi-symmetric structure, while younger ones (5-7 Gyr) trace the main Galactic bar. Similarly, 
\citet{iwanek+23} found that bulge Miras trace well the Galactic bar and also follow the B/P structure. In VVV, on the other hand, most Mira 
variables are saturated, except for those in the high extinction region close to the midplane
\citep{sanders+22mirasNSD}. The latter study found that the age distribution of Miras in the nuclear stellar disk is relatively flat between 2 
and 10 Gyr, though with large uncertainties due to the spread of the period-age relation, especially towards the young (long period) side. 
These authors also emphasize that they cannot distiguish between the population of the nuclear stellar disk and that of the underlying bulge. 
In other words, their variables belong to the central, high extinction region, not necessarily to a distinct component identifiable as a 
nuclear stellar disk.

\subsubsection{Classical Cepheids}   
These bright variables have been reported absent, in the Galactic bulge and in general within the inner 2.5 kpc of the Galactic disk 
\citep{matsunaga+16}. This is not surprising, as this class of variables traces stellar populations in the age range 10-300 Myr \citep{bono+05}. 
Only four of them have been found close to the Galactic center, claimed to belong to the nuclear stellar disk \citep{matsunaga+13, matsunaga+15}.

\subsection{The metal poor spheroid}   
\begin{figure}[h]
\includegraphics[width=\textwidth]{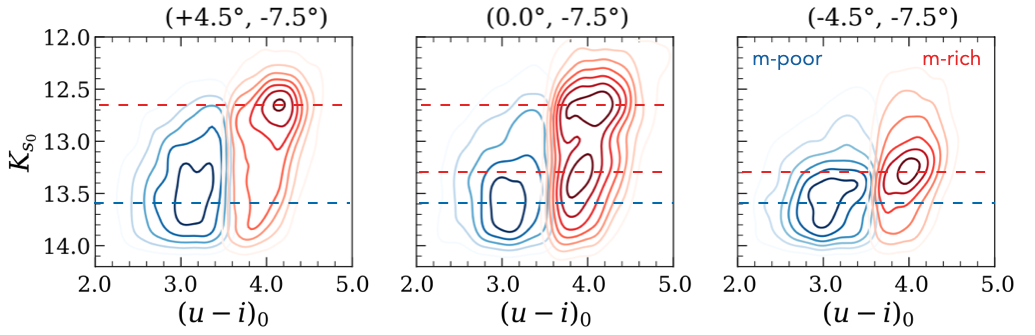}
\caption{Density contours distribution of RC stars in the [K$_s$, ($u-i$)$_0$] CMD plane for three different fields at b=$-$7.5$^\circ$ and longitudes $\ell$=+4.5$^\circ$ (left), $\ell$=0$^\circ$ (middle) and $\ell$=$-$4.5$^\circ$ (right). The RC population is divided in two subgroups according to their ($u-i$)$_0$ color, which is a proxy for the metallicity. Figure adapted from \citet{lim+21}.}
\label{fig:2RC}
\end{figure}

Soon after the B/P or X-shape was discovered through the presence of a double RC in the lines of sight at $|\ell|\sim$0$^\circ$ and $|b|>$5$^\circ$, 
a few studies claimed that this RC split was not present in the CMD obtained by selecting only m\--poor stars \citep{uttenthaler+12, ness+2012_2RCs, 
rojas-arriagada+14}. This was a first indication of what was later established beyond any doubts: m\--poor stars in the MW bulge are not arranged 
in either a bar nor a B/P: they trace a spheroidal distribution (see also Fig. \ref{fig:rot-mpmr}a). 
Spectroscopic analysis of bulge RC stars revealed that their iron distribution is everywhere bimodal, with a gap at [Fe/H]$\sim$0 dex. Further, 
only the m\--rich stars follow the bar, while the m\--poor ones trace an axisymmetric structure \citep{zoccali+2017}. The BDBS survey confirmed 
and reinforced the same results, by means of photometric metallicities of 2,6 millions RC stars \citep{johnson+22}. \citet{johnson+20}
showed that the ($u-i$)$_0$ color of RC stars nicely correlates with [Fe/H] from high resolution spectroscopy. A cut at ($u-i$)$_0$=3.5 separates
m-poor from m-rich RC stars, similar to a cut at [Fe/H]$\sim$0. Figure~\ref{fig:2RC}, from \citet{lim+21}, shows that the magnitude of the m-poor 
RC does not vary with longitude. On the contrary, the magnitude of m-rich RCs shows the known variation due to the presence of the bar, together
with the splitting at $\ell$=0$^\circ$ due to the X-shape. This very direct result,
not involving complicated maps nor spectroscopic measurements, raises an important red flag upon the bar parameters derived from RC stars, discussed above. Indeed, all of them used RC stars without distinction on their metallicity. As a consequence, those works fitted a single bar+B/P component on tracers that are, in fact, a mix of two populations. To date, the only large catalog of RC stars that includes metallicity is the one from BDBS \citep{johnson+22}, that is limited to relatively large, negative, distances from the midplane (b$\lesssim$$-$4$^\circ$, i.e., Z$\sim$600 pc), because it comes from an optical survey. The absence of the, more populated, inner region has so far discouraged new derivations of the bulge shape parameters as a function of metallicity.


\section{The age of stars in the Galactic bulge}\label{sec:age}

As we have seen in previous sections, to understand what is the bulge one has to look at its stellar population properties, such as the age, the chemical composition and kinematics: "the three dominants parameters in the population concept" (cfr., Sandage).
However, among these three parameters, the age becomes particularly crucial when the focus is shifted to a more complex question, i.e., {\it How did the Bulge form?}
Indeed, various proposed formation processes (e.g., dynamical instability, hierarchical mergers, gravitational collapse) are expected to take place on different timescales, hence the related models predict different ages for the bulge stellar population.

Historically, the first attempt to infer the age of the bulge comes from the simple, but effective, {\it by association} indirect method that relies on detecting bulge sources  unequivocally tracing either young, MW disk-like (called Pop\,I) or old, GC-like (Pop\,II) stars. 
In 1951 and 1958, Baade detected RRL variables in a low extinction bulge field at $(l, b)=(0^\circ, -3.9^\circ)$, a region that 
has since taken the name of Baade's Window (BW), becoming subject of extensive studies over many decades. 
Being m-poor and old, RRL variables were already recognized as tracer of Pop\,II, thus Baade concluded that the bulge was a population similar to those of GCs. However, the presence of RRL variables did not exclude the presence of a spread in age nor in metallicity.
While Baade's conclusion on the low metallicity of bulge stars was challenged shortly after by several photometric and spectroscopic studies \citep[see ][ and section \ref{sec:chem}]{Frogel88}, his claim about the old age was instead further reinforced by the non-detection of any luminous carbon stars (tracers of intermediate age populations) in the survey, based on near\--IR low\--resolution spectra of M giants, in the BW and another two fields at $b=-8^\circ, -10^\circ$\citep{Blanco+84,Blanco+86}. 
The presence of GCs in the bulge (e.g., NGC\,6553 and NGC\,6528), known already back then, was yet another evidence that 
the bulge is hosting an old population. However, it was only with the first studies of the MS\--TO in the observed CMD of several bulge fields that direct age estimates for the bulk of the population were finally possible.

Being {\it the core of the stellar evolution  clock} \citep{RenziniFusiPecci88}, the luminosity of the MS\--TO is the best  diagnostic to infer the age of both simple and complex stellar populations. Indeed, isochrones for different age and metallicity are clearly separated at the MS\--TO level, as opposed to other sequences such as the RGB and the RC, which are instead more sensitive to the metallicity and He content, respectively. Thanks to the weak sensitivity of the RC to age, discussed in Sec.~\ref{sec:3D}, the observed magnitude difference between the MS\--TO and the RC in a CMD provides a handy and powerful tool to estimate the age of a given population once compared with theoretical expectations.

\begin{figure}[b]
\centering
\includegraphics[width=\textwidth]{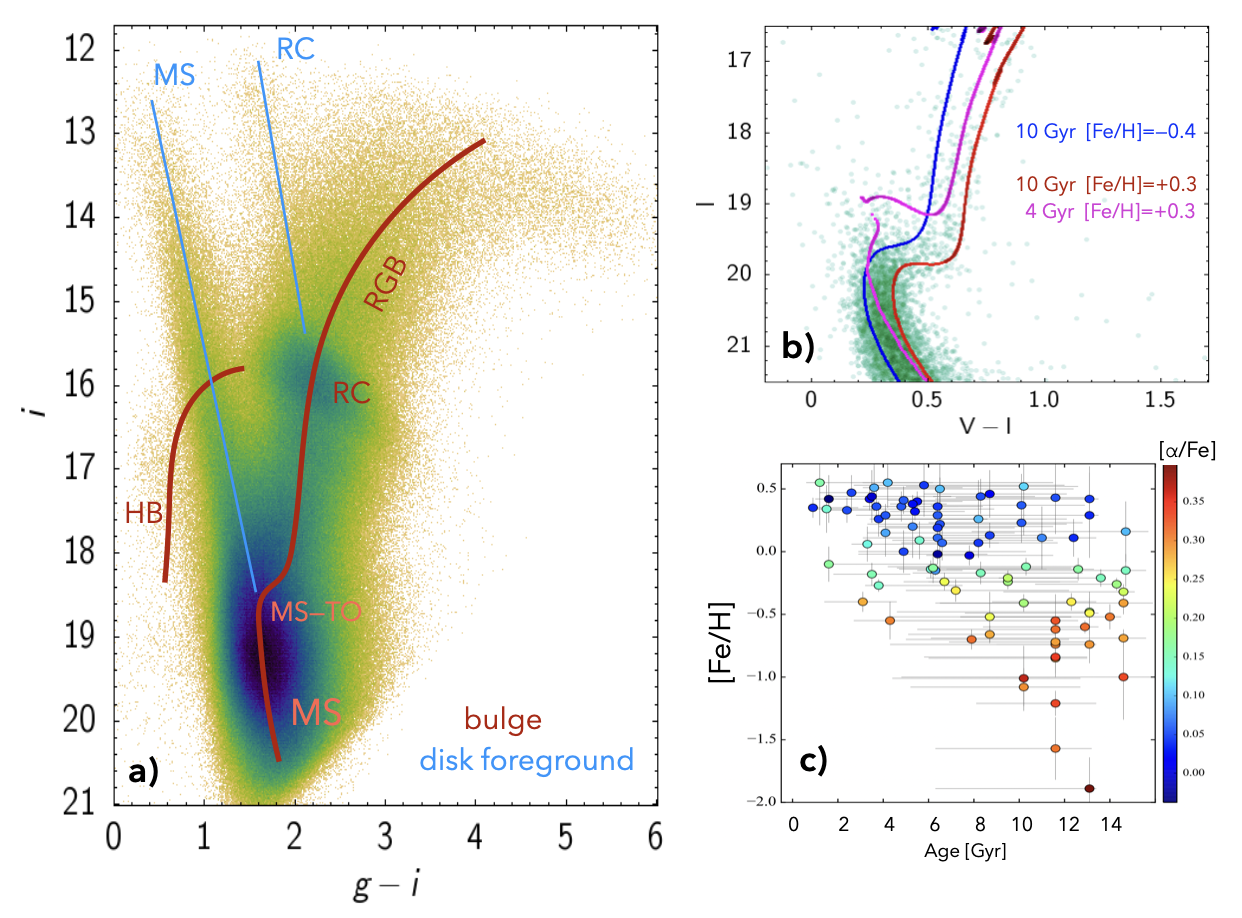}
\caption{{\bf Panel a)}: Observed CMD of the BW from \cite{Saha+19DECam}. Labels and lines mark key evolutionary sequences discussed in the text. \-- {\bf Panel b)}: Isochrones\--fitting of the HST\--based CMD of a bulge field where the contribution from the foreground disk has been previously removed by using stellar proper motions \citep[Figure adapted from ][]{clarkson+11}. \-- {\bf Panel c):} Age\--metallicity relation as obtained from 90 microlensed bulge dwarfs \citep[Figure from ][]{bensby+2017}.}
\label{fig:AGE}
\end{figure}

In practice, dating bulge stars is a very tricky task, challenged by several observational issues, such as the presence of a spread in distance and interstellar extinction, both broadening the observed sequences, and the presence of contaminating disk foreground stars.
Specifically, as shown in Fig.\,\ref{fig:AGE}a, the disk main sequence, spread along a large magnitude (distance) range, heavily contaminates precisely the bulge MS-TO region, hampering an accurate analysis of the main age-dependent feature of its CMD.

The first attempt to detect the MS\--TO in a bulge field at b$=-8^\circ$ dates back to \cite{vandenBerg74_MSTO} by using photographic plates. 
The derived luminosity function obtained from star counts was found consistent with GC\--like age. However, a more solid and convincing 
detection was presented by \cite{terndrup88} thanks to deep CCD based photometry, which allowed to construct the very first CMD of an outer
bulge region (b$=-10^\circ$) sampling the population from the RC down to $\sim$1\,mag below the MS\--TO. 
From the comparison between the CMD and theoretical isochrones, Turndrup derived an age of 11\--14 Gyr for the bulge population, ruling out
a substantial presence of  any population younger than $\sim$5\,Gyr. 
In addition, by using the Galactic model of Bahcall and Soneira \citep{BahcallSoneira80,BahcallSoneira84,Bahcall86rev}, \cite{terndrup88} 
was the first to provide convincing evidence that the blue vertical  sequence, present in the CMD of all observed bulge fields, belongs to
the foreground disk.

Over the subsequent decades, many studies of the MS\--TO from  space\-- \citep[i.e., ][]{ortolani+95,FeltzingGilmore00,kuijken+02,clarkson+08,clarkson+11} and ground\--based photometry \citep[i.e.,][]{zoccali+03,valenti+13,surot+19age}
in different bulge fields, have all contributed to consolidate the consensus that the bulge hosts an old (i.e., $\geq10$\,Gyr)
population with negligible contribution from young ($\leq5$\,Gyr) stars. 
In particular, \cite{clarkson+08} provided the first space based photometry kinematically decontaminated from foreground disk, by means of 
stellar proper motions. The resulting CMD, shown in Fig.~\ref{fig:AGE}b, shows that most stars in the MS-TO region are compatible with a
purely old population, with a spread in metallicity. According to \citet{clarkson+11}, who analyzed their light curves, the stars brighter and bluer than the main MS-TO are likely blue straggler stars (i.e., stars as old as the fainter ones, but more massive because of mass transfer from a binary companion). 
The same authors set an upper limit of $\sim3.4\%$ to a possible bulge component younger than 5\,Gyr. It is worth mentioning that the 
agreement and consistency among the above mentioned studies is really remarkable when considering the differences in the adopted 
instrumentation (i.e., space vs ground), wavelength coverage (i.e., optical vs near\--IR) and disk-decontamination method (i.e., 
kinematics vs statistic). 

Nevertheless, there is a clear discrepancy between the ages inferred from the MS\--TO in the observed CMD and those derived 
spectroscopically. The latter have been measured only recently, thanks to the availability of comprehensive spectroscopic bulge 
surveys such as APOGEE. The results from the microlensing project led by Bensby and collaborators are pivotal in the context of 
the long\--standing debate on bulge ages. 
Specifically, \cite{bensby+13,bensby+2017} observed 90 F and G bulge dwarf, MS\--TO and subgiant stars in the bulge (i.e., {l}$\leq6^\circ$
and $-6^\circ \le b \le 1^\circ$) during microlensing events, which amplify the light of the otherwise too faint dwarfs, thus enabling 
high\--resolution (R$\ge$80,000) spectroscopic measurements. Such high\--resolution spectra enabled the Bensby team to derive individual 
stellar ages from the effective temperature and gravity (i.e., from isochrones in the [T$_{eff}$, log\,g] plane).

They found that while the vast majority of the sampled m-poor stars ([Fe/H]$\lesssim$$-$0.5\,dex) are 10\,Gyr or older, about 35\% of the 
observed m-rich stars (Fe/H]$\ge$0\,dex) span ages in between 2\,Gyr and 8\,Gyr. 
In addition, from the so\--derived age\--metallicity and age\--$\alpha$ elements distribution (see Fig.\,\ref{fig:AGE}c), they concluded 
that the bulge must have experienced several significant star formation episodes, about 3, 6, 8 and 12\,Gyr ago. 
Comparable results have been found by \cite{Schultheis+17} and later by \cite{Hasselquist+20}, who derived the age distribution of bulge
giants from APOGEE spectra, by using the [C/N]\--[Fe/H]\--age relation, calibrated on asteroseismic data. In both studies, the m-rich 
sample has been found to include stars as young as 1\--2\,Gyr.

Different concepts have been proposed to reconcile, at least partially, the spectroscopic and photometric ages. In this respect, the 
first attempt was presented by \citet{Nataf+12} and \citet{Nataf16}, arguing for the possibility that the bulge m-rich population might 
have a He overabundance with respect to the standard value. The use of standard isochrones on He\--enriched population would lead to 
photometric and spectroscopic ages that are over\-- and under\--estimated, respectively. Therefore, if the bulge chemical evolution were 
to produce excess He, which is still to be proven and quantified, then the discrepancy would be substantially reduced.

An alternative approach, as opposed to fitting the MS\--TO of the CMD with a set of isochrones, has been proposed by \cite{Haywood+16} and 
\cite{bernard+2018}. These authors used synthetic CMDs to match the observed ones, thus reconstructing the bulge star formation history. 
Attempting to reproduce the HST\--based CMD of \cite{clarkson+08}, both studies advocated for the presence of multiple star formation 
episodes (i.e., similar to Bensby's results). In this context, it is fair to mention that the synthetic CMD approach is in principle superior
to the isochrone fitting one, because it allows us to model both the observational errors and the intrinsic spreads due to distance, 
extinction, metallicity and age. Nonetheless, the accuracy of the result critically depends on the detailed modeling of these interconnected and often degenerate quantities. In this sense, in order to break the degeneracies it is critical that the model can simultaneously reproduce
all the observed features of the CMD. 

Certainly, problems exist with all dating methods. Specifically, in the microlensing approach, individual metallicities are well measured, 
but the method is more prone to the temperatures of the chosen isochrones, which are less robustly predicted by the stellar evolution theory
compared to MS\--TO luminosities, on which the CMD approach is instead based. Indeed, by using different set of isochrones with the same 
\cite{bensby+2017} stellar quantities, \cite{Joyce+23} concluded that only a few percent of bulge stars would be as young as 1-2\,Gyr. 
Additionally, the method relying on the [C/N]\--age relation is also limited by the uncertainties in the individual stars distance, which 
can be as large as 20\--30\% (e.g., stars attributed to the bulge could well be located in the disk instead).
The isochrone\--fitting approach is affected by the age\--metallicity degeneracy. That is, not knowing the metallicity from other sources,
the position in the CMD of a given MS\--TO star can be fitted equally well by a m-rich, young model or by a m-poor old one. Decontamination
from foreground disk stars is also critical. In an attempt to break the age\--metallicity degeneracy, \cite{Brown+10} designed a combination
of optical and near\--IR HST filters defining two reddening free indices sensitive to metallicity and temperature, that have been used to 
map different bulge fields.
Such photometric database has been used by \cite{renzini+2018} to perform the metallicity tagging of near MS\--TO stars. They found that 
the m-poor and m-rich populations have consistent luminosity functions around the MS\--TO, hence appearing coeval and $\sim$10\,Gyr old, 
with at most few percent of stars as young as $\sim$5\,Gyr.

Finally, by combining metallicities and radial velocities from APOGEE, proper motions from Gaia\,DR3 and distances from StarHorse, 
\citet{queiroz+20} and \citet{queiroz+21} mapped gradients of chemical abundances  and ages for $\sim$8000 giants within 5\,kpc 
from the Galactic centre (see Sec.~\ref{sec:alphas}). They suggest that the inconsistency between the photometric and spectroscopic 
ages could be due to different selection functions. While the photometric studies have, so far, mostly sampled the old, 
pressure\--supported spheroidal component within 2\--3\,kpc, the spectroscopic studies \-- especially those using big surveys \-- 
deal with samples that consist of mixed population belonging to the bar and thin disk. In particular, they ascribe the very young 
ages present in the spectroscopic age\--metallicity distributions to the contribution of the thin disc.

To summarize, nearly 60 years after the first estimate by Baade, the age of the bulge stars is still a controversial and debated topic. 
While there is a consensus on an old (i.e., $\geq$10\,Gyr) age for the bulk of the bulge population, especially for the m-poor component, the question regarding the possible presence of intermediate\--young stars (i.e., $\leq$5\,Gyr), and in what fraction, remains open to these days.

\section{The bulge IMF}\label{sec:IMF}

The evolutionary path and lifetime of any given star is primarily driven by its mass. Therefore, the number of stars born at any given
mass, known as the Initial Mass Function (IMF), is critical to the formation and evolution of any stellar system. Because the small 
number of massive stars dominate the light of a system, while the large number of low-mass stars dominate the mass, the IMF drive the
luminosity evolution over time, and mass\--to\--light ratio (M/L). Further, only massive stars explode as supernovae, their relative 
fraction regulates both the chemical enrichment and the energetic feedback. 

Although the age distribution of the bulge m-rich component is still controversial, the bulge present\--day mass function below the 
old MS\--TO (i.e., $M\leq1\,M_{\odot}$) is the same as the IMF, because those are unevolved stars. In addition, because the bulge
relaxation time is larger than its age, dynamical processes did not have time to segregate stars of different masses. Therefore, the 
bulge IMF can be derived directly from the observed luminosity function of MS stars, which can be converted into a mass function by
means of a mass\--luminosity relation.

To these days, the bulge IMF, in the low\--mass range, has been the subject of only three HST\--based studies: in the BW by 
\citet[][ for $1\,M_{\odot}\lesssim M \lesssim 0.3\,M_{\odot}$ ]{holtzman+1998}, in a field along the bulge minor axis at 
$b=-6^\circ$ by \citet[][ for $1\,M_{\odot}\lesssim M \lesssim 0.15\,M_{\odot}$ ]{zoccali+00}, and in the innermost field at 
$b\sim-2.7^\circ$ by \citet[][ for $1\,M_{\odot}\lesssim M \lesssim 0.15\,M_{\odot}$ ]{calamida+15}.
These studies followed different prescriptions in terms of adopted mass\--luminosity relation, foreground disk decontamination, 
and unresolved binary treatment. However, within the uncertainties, their results are remarkably consistent, pointing towards a 
power-law IMF ($\Phi$(M)$\propto$M$^\alpha$) with exponent $\alpha$ smaller than the one originally proposed by Salpeter 
($\alpha$=$-$2.35), and thus more consistent with a Kroupa's IMF \citep{kroupa02}.
The work by \cite{calamida+15} currently represents the state-of-the-art, given the treatment of unresolved binaries and the use 
of proper motions to derive the IMF from a clean bulge sample. They found that the bulge IMF is best reproduced by a two power 
low functions with a break at $M\sim 0.56\,M_{\odot}$ and slopes $\alpha=-1.41$ for masses between $1\,M_{\odot}\leq M \leq0.56\,M_{\odot}$, and $\alpha=-1.25$ in the range $0.56\,M_{\odot}\leq M \leq0.15\,M_{\odot}$. Broadly consistent results are 
found by \cite{Kirkpatrick+2024} for the IMF of the nearby disk based on Gaia and Spitzer data. Finally, the bulge IMF appears 
consistent with that of massive Galactic GCs, far enough from the Galactic plane, so that they are not significantly affected 
by dynamical processes. This reinforces the consensus that the chemical composition of a collapsing cloud has little impact on 
the resulting IMF.

\section{The bulge total mass}\label{sec:mass}

Because the mass is likely the main parameter governing galaxy evolution, several studies focused on the MW dynamical mass
(i.e., baryonic + dark). Historically, this was done by using dynamical models to match either the galactic rotation curve
inside $\sim$1\,kpc, or the kinematics (i.e., velocity and velocity dispersion) of different stellar tracers. Then, by 
assuming an observed luminosity profile (often in K\--band) the M/L ratio can be derived, and hence the final bulge mass.
Over the last few decades, different studies have provided bulge mass values spanning more than a factor of two, from 
3$\times$10$^{10}$\,M$_{\odot}$, \citep{SellwoodSanders88} to 0.6$\times$10$^{10}$\,M$_{\odot}$, \citep{Sofue13}. 
The M/L ratio derived from fitting the rotation curve has been found to be consistently a factor of 2-3 larger than that
obtained from stellar kinematics. Such large discrepancy in the M/L determination has been often ascribed to the presence 
of large non\--circular motions distorting the rotation curve and thus leading to an overestimation of the M/L ratio.
Additionally, as pointed out by \citet{BlandHawthornGerhard16}, for a barred bulge like ours, hydrodynamical models are 
needed to properly analyse the full gas velocity field as opposed to simple rotation curve analysis.
When considering only the studies based on stellar kinematics, the resulting bulge dynamical estimates cluster around 
1.8\--2$\times$10$^{10}$\,M$_{\odot}$. A few studies found values as large as 2.8$\times$10$^{10}$\,M$_{\odot}$ \citep{Blum95} or 
as small as 1.2$\times$10$^{10}$\,M$_{\odot}$ \citep{Kent+91,Kent92}.
In this context, the most recent dynamical mass determination is provided by \citet{Portail+15} using a made\--to\--measured dynamical
model of the bulge, with different dark matter halos, to match the stellar kinematics from the BRAVA survey
\citep{Rich+07BRAVA,kunder+2012} and the 3D surface brightness profile derived by \citet{wegg+13} from VVV data.
Their best\--fit model is consistent with the bulge having a dynamical mass of 1.8$\pm$0.07$\times$10$^{10}$\,M$_{\odot}$.

Finally, for the sake of thoroughness, the bulge dynamical mass distribution has been also constrained using independent methods such
as the frequency of microlensing events \citep[e.g.,][]{mroz+19, kaczmarek+24}, and with gravitational waves \citep[e.g.,][for the expected determination based on LISA mission]{wilhelm+21}. 
These works demonstrated the potential of the methods, although the results are not yet competitive with those based on kinematics of
stellar tracers, due to the uncertainties associated to the frequency and rate of the events.

Assumptions on the dark matter content in the bulge allowed the stellar mass to be derived from the dynamical mass. If the contribution
of the dark matter to the bulge is negligible, then dynamical mass should corresponds to the stellar mass. However, the stellar mass has
been also directly constrained by using stellar population models to match either surface photometric profiles (e.g., COBE) or star
counts (e.g., DENIS, VVV, OGLE photometric surveys). Also for the stellar mass case, the quoted values in the literature span a fairly
large range, between $1.3 \times 10^{10}M_{\odot}$ \citep{dwek+1995} and $2.4 \times 10^{10}M_{\odot}$ \citep{picaud+04}.
\citet{valenti+2016} scaled the observed VVV RC stellar density map with the observed luminosity function from \citet{zoccali+00} and
\citet{zoccali+03}, thus providing the first empirical (i.e., no model\--dependent) estimate of the bulge stellar mass. From the 
observed mass profile in the bulge region $|b|\le$9.5$^\circ$ and $|l|\le$10$^\circ$, they inferred a mass in stars and remnants of 
2$\pm$0.3$\times$10$^{10}$\,M$_{\odot}$.

To summarize, the most recent estimates of the bulge dynamical \citep{Portail+15} and stellar \citep{valenti+2016} mass are compatible
within the quoted errors. The consistency between these findings improves when considering that: {\it i}) the empirical estimate of the
stellar mass by \citet{valenti+2016} refers to a larger volume that is not limited along the line of sight, and {\it ii}) depending on
the IMF adopted in the model of \citet{Portail+15}, about 10\--40\% of the bulge mass should be in the form of dark matter thus leading
to a stellar mass between 1.4\--1.7$\times$10$^{10}$\,M$_{\odot}$.




\section{Bulge Kinematics}\label{sec:kine}

\begin{figure}[t]
\centering
\includegraphics[width=0.8\textwidth]{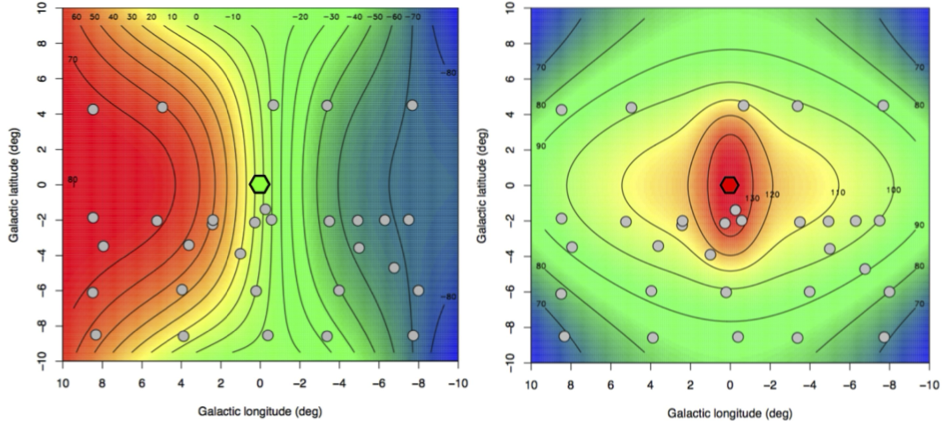}
\caption{Rotation (a) and velocity dispersion (b) maps of the MW bulge. Grey circles mark the location of the fields targeted by the GIBS spectroscopic survey and used to construct the maps. The big hexagon refers to the mean position of 4 fields observed by \citet{valenti+2018} to verify the adequacy of the maps in a bulge region otherwise poorly constrained by the GIBS survey. Figure adapted from \citet{zoccali+14}.}
\label{fig:rot}
\end{figure}

Large spectroscopic surveys such as BRAVA \citep[][]{kunder+2012}, Gaia ESO \citep{rojas-arriagada+14}, and ARGOS 
\citep{ness+2013kin_argosIV} consistently found that bulge stars rotate with almost cylindrical rotation (Fig.~\ref{fig:rot}
a). The latter has been defined for the first time by \citet{kormendy+12}, as a rotation with velocity that is almost 
constant with height above the plane. Based on this evidence, some authors argued that the MW is a pure disk galaxy, i.e., 
its kinematics is consistent with that of models with little or none pressure-supported (spheroidal) component, in the
center \citep{howard+2009, debattista+2017}. More recent data, including stars closer to the midplane, showed that the
innermost region (R$\lesssim$150 pc) shows a peak in velocity dispersion that might require another component 
\citep[Fig.~\ref{fig:rot}b;][]{zoccali+14, valenti+2018}. 

The maps in Fig.~\ref{fig:rot} represent the mean velocity of stars along the line of sight, with respect to a reference frame where 
the Galactic center is at rest, and the motion of the Sun, relative to each star, has been subtracted. As such, they allow to 
compare the MW bulge with external galaxies observed with Integral 
Field Units. Indeed, by comparing the kinematics of our bulge with that of 12 disk galaxies with mid to high inclinations ($\sim$ 
edge on) \citet{molaeinezhad+16} found that the MW bulge has a 72$\%$ cylindrical rotation, thus comparable to that of the strongest 
barred galaxy in their sample. 

By analyzing maps of proper motions from the VIRAC catalog \citep{virac}, \citet{sanders+19barps} and \citet{clarke+19barps} proved
that the differential rotation between the two RCs seen along the bulge minor axis confirms the B/P/X-shaped nature of the bulge, 
ruling out the alternative explanation that the split RC was due to the presence of two populations with different chemical compositions
\citep{lee+15,gonzalez+15}.

Superimposed to the orbits of individual stars, the MW bar rotates as a solid structure, with a pattern speed that has been measured 
to be very close to $\Omega_{\rm P}$=40 km/s, with a notable consensus \citep[e.g.,][]{portail+17barps, shen+20, sanders+19barps, clarke+19barps, li+22, lucchini+24}.

The few studies with large enough statistics to address the question whether the m\--poor stars rotate slower than the m\--rich ones
have all found a difference in this sense, but its magnitude is not large \citep{kunder+2016, clarkson+2018, ara+20}. 
Concerning the velocity dispersion, the difference is more clear, with m\--poor stars showing a large dispersion (100-120 km/s), roughly constant across the plane of the sky, while m\--rich stars have a marked gradient, ranging from $\sigma_{\rm RV}$=60 km/s at 
b=$-$8$^\circ$ and $\sigma_{\rm RV}$=140 km/s at b=$-$1$^\circ$ \citep{zoccali+2017, valenti+2018}. 

A measurable kinematical quantity that is strongly affected by the presence of the bar, and its inclination, is the so-called {\it 
vertex deviation}, defined as the angle between the major axis of the velocity ellipsoid (in a radial versus tangential velocity 
plane) and the line of sight direction. In the absence of a bar, a pressure supported axisymmetric spheroid would have vertex deviation
equal zero. The amount and sign of a non zero vertex deviation, instead, depends on the bar inclination angle. This thus represents an
independent, purely kinematical way to constrain both the presence of a bar and its angle. \citet{babusiaux+10} showed that, in the 
Galactic bulge, the vertex deviation is large and negative for m\--rich stars, while it goes to zero for [Fe/H]$<$$-$0.5. This was an
early evidence for the m\--poor stars not following the Galactic bar.  More recently, \citet{simion+21} calculated the variation, across
the sky, of the vertex deviation for a sample of $\sim$7000 stars with 3D velocities and abundances from the ARGOS and the Gaia DR2
surveys. Their results is less clear, probably due to the fact that, by separating the stars in 17 positions across the sky and 3
metallicity bins at each position, they end up with relatively low statistics. Broadly, though, they confirm that the vertex deviation
is zero at b=$-$10$^\circ$, where the m\--poor population dominates, while it is large and negative at b=$-$5$^\circ$, where the 
m\--rich population is more prominent. They constrain the inclination angle of the Galactic bar to be 29$\pm$3 degrees, consistent 
with RC maps.

\section{Iron and $\alpha$-elements surface abundance of bulge stars}\label{sec:chem}

Constraining the shape and peak of the MDF of any stellar system provides crucial insights on the IMF, star formation efficiency and
possible gas infall timescale \citep{matteucci+99}. On the other hand, to unveil the chemical evolution enrichment of a system, one 
should look at the abundances and abundances pattern of several key elements such as iron\--peak, CNO and $\alpha$\--elements. The 
latter are synthesized from $\alpha$ particles (i.e., He nuclei) such as O, Ne, Mg, Si, Ca and S.

Indeed, chemical elements observed in the stellar atmospheres are synthesized in stars of different masses, hence released into the
interstellar medium on different timescales. In this context, the study of the [$\alpha$/Fe] vs [Fe/H] trend is particularly handy
because it is driven by the time delay in the production of the bulk of Fe and Fe\--peak elements (ie., SNe\,Ia, low mass progenitors),
relative to $\alpha$\--elements (i.e., SNe\,II, high mass progenitors). Therefore, it provides a sort of chemical clock, whereby high
[$\alpha$/Fe] populations were born rapidly, within a high star formation environment, while low [$\alpha$/Fe] populations were formed 
at a slower pace.

\begin{figure}[b]
\centering
\includegraphics[width=0.7\textwidth]{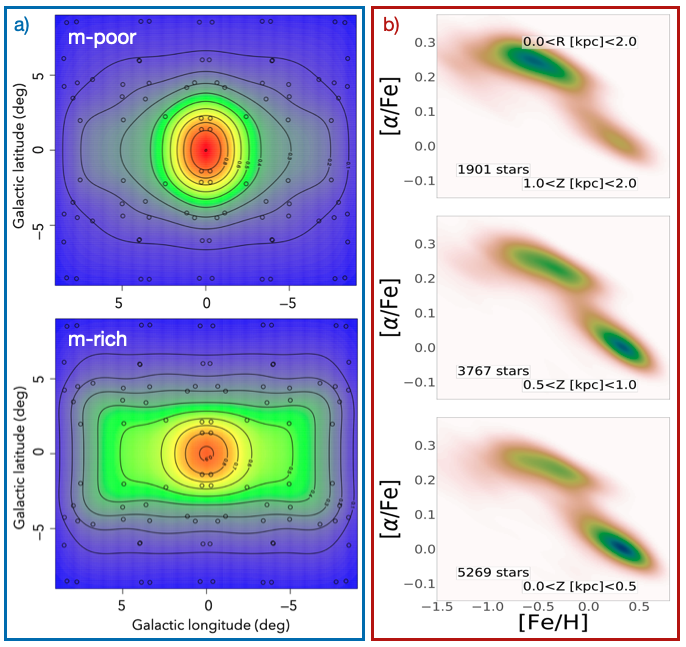}
\caption{
Panels a): Density map of m-poor (top) and m-rich (bottom) RC stars obtained using the GIBS sample and the star counts profile from \citet{valenti+2016}. Figure adapted from \citet{zoccali+2018}.
Panels b): APOGEE [$\alpha$/Fe] vs [Fe/H] distribution in bins of galactocentric cylindrical coordinates. The distance from the Galactic plane increases along the panels column (bottom to top), while the distance from the Galactic center increases along the row panels (left to right). Figure adapted from \citet{queiroz+20}.
}
\label{fig:rot-mpmr}
\end{figure}

\subsection{The metallicity distribution function}\label{sec:mdf}
Early indications on the bulge metallicity, and the possible presence of a gradient, have been reported by \citet{rodgers+86_MSTO} and 
\citet{terndrup88} using the photometric color of RGB stars as a proxy for the metallicity, in the CMD of different fields.
However, we owe the very first MDF based on high\--resolution spectroscopy to the pioneering work of \citet{mwr+1994}, who observed 
11 K giants in the BW. Shortly after, many other studies \citep[e.g.,][]{Fulbright+06,RichOri07,zoccali+2008,johnson+11,gonzalez+11spec,hill+11,rich+12} targeted a handful of sparse fields, mostly located
along the bulge minor axis. A remarkable consistency was found in the resulting MDFs, all showing the bulge population to span a 
rather broad metallicity range, $-$1.5$\lesssim$\,[Fe/H]\,$\lesssim+0.5$\,dex, with a peak at solar or even super-solar metallicity.
Since then, our understanding of the metallicity of the bulge population has improved tremendously, thanks to recent spectroscopic
surveys such as ARGOS, BRAVA, GIBS, Gaia ESO and APOGEE. By targeting thousands of K and M giants from the bulge outskirts to the
innermost regions, they enabled the determination of the MDF across most of the bulge extension. The overall emerging picture 
depicted by the surveys clearly highlights the presence of two populations: a m-rich and a m-poor component \citep[e.g.,][]{ness+2013mdf_argosIII,zoccali+14,rojas-arriagada+2014}.
Indeed, nearly anywhere across the bulge the observed MDF is bimodal, with a m\--poor peak around [Fe/H]=$\sim$$-$0.3\,dex and a m-rich
one in the super\--solar regime (at [Fe/H]$\sim$+0.3\,dex). The observed mild negative gradient of the {\it mean} metallicity versus
latitude, is actually due to the combination of these two components, whose relative fraction changes across the bulge. 
As shown in Fig.\,\ref{fig:rot-mpmr}a, these two components have very different spatial distribution: the m-poor component has a spheroid\--like shape, versus a boxy/bar distribution of the m\--rich component \citep{zoccali+2018}. 
Such difference is independently confirmed by the observed magnitude distribution of RC stars, shown in Fig.~\ref{fig:2RC} \citep[see also:][]{ness+2013mdf_argosIII,rojas-arriagada+2014,rojas-arriagada+2017}. 
Recent N\--body simulations of an evolving disk galaxy show that the observed spatial distribution of these two populations can
be successfully reproduced as the result of dynamical evolution of two disks with different initial velocity dispersions, similar to
the MW thick and thin disk \citep[e.g.,][]{debattista+2017,fragkoudi+2018}. In other words, the observed m\--poor spheroid might have 
formed in-situ, from the thick disk. Nonetheless, at present we cannot state that this scenario is to be preferred, compared to
another one where the spheroid is the result of accretion processes occurring before the formation of the main bar \citep[e.g.,][]{athanassoula+17}.

As expected, the bulge MDF traced by K and M giants is remarkably different from that observed when using RRL. 
Indeed, only stars in the bulge m\--poor tail burn He in their core at sufficiently blue colors to cross the instability strip and thus
pulsate. Although less accurate, as based on photometric estimates, the MDF of the RRL population across the bulge is broader, extending 
to much lower metallicity (i.e., $\sim$$-$2.5\,dex), and it shows a single peak. On the other hand, both the location of the peak and the
m\--rich end differ substantially among the recent studies. The MDF based on OGLE\--IV catalogs, in the region between $|l|\lesssim10^\circ$
and 2$^\circ$$\lesssim$|b|$\lesssim$8$^\circ$ peaks at [Fe/H]$\sim$$-$1\,dex and it extends to $\sim$+0.5\,dex \citep{pietrukowicz+15}.
Conversely, the RRL in the inner bulge region, within $|l|$$\leq$10$^\circ$ and $|b|$$\leq$+2.5$^\circ$, is more m\--poor ($-2.5\leq$ 
[Fe/H] $\leq -0.5$), with a peak around $-$1.5\,dex \citep{zoccali+24RRL}. Before interpreting such discrepancy as evidence of a metallicity 
gradient, with the RRL in the innermost region being more m-poor, the possible systematics and errors associated to the photometric
metallicity estimates must be investigated. 
This is particularly true when considering that the only spectroscopic MDF of bulge RRL, obtained by measuring CaT lines on the BRAVA
spectra \citep{Savino+2020} spans a very large range $-2.5\lesssim$[Fe/H]$\lesssim+0.5$ \citep[i.e., consistent to][]{pietrukowicz+15}, 
and it shows a single peak at [Fe/H]$\sim$$-$1.4\,dex \citep[i.e., similar to][]{zoccali+24RRL}.
Despite the above mentioned differences, overall the MDF of bulge RRL appears to be narrower and more m\--rich than that observed for halo
RRL stars \citep[see e.g.,][]{CabreraGarcia+2023}.

\subsection{The $\alpha$\--elements distribution}\label{sec:alphas}

The determination of $\alpha$\--elements abundance with sufficient accuracy (i.e., $\leq$0.2\,dex) necessarily requires high\--resolution 
(i.e., R$\gtrsim$ 18,000) spectra. Because they are expensive, early 
studies on $\alpha$\--elements abundance were limited to small sample of giants in few bulge regions, along the minor axis 
\citep[see review by][and reference therein]{mcwilliam+2016}. These studies showed a general trend whereby bulge stars are $\alpha$\--enhanced 
by $\sim0.2$\--0.3\,dex, with respect to the solar value, in the m-poor regime, and then progressively $\alpha$\--poor in the super\--solar 
regime. This trend is remarkably similar to that of thick disk stars \citep[e.g.,][]{Bensby+11disk}, although they do 
not reach the same high metallicity range seen in the bulge. As it has been the case for the MDF, the advent of the spectroscopic surveys has 
enabled us to confirm and expand the earlier studies with unprecedented details 
\citep[e.g.,][]{ness+2013mdf_argosIII,gonzalez+15gibsII,rojas-arriagada+2017}. Particularly, the bimodality already seen in the MDF became evident
also in the [$\alpha$/Fe] trend \citep{ara+19}. 

Most notably, \cite{queiroz+20} clearly identified a m\--poor, $\alpha$\--rich, population dominant at large distances from the midplane, and
another m\--rich, $\alpha$\--poor one dominant close to it (Fig.\,\ref{fig:rot-mpmr}b). The observed [$\alpha$/Fe] vs [Fe/H] trend unequivocally defines a true chemical discontinuity, strongly supporting a different formation timescale.

While we still lack a comprehensive understanding of the origin of the observed bimodality, an interesting scenario was proposed by 
\citet{Debattista+23}. They perform N\--body hydrodynamics simulations of a high\--redshift disk galaxy, successfully reproducing the
presence of high star formation clumps, observed in high redshift disks \citep{forster-schreiber20} by suppressing feedback. At the
present time, bulge stars that originally formed within one of the disk clumps are $\alpha$\--rich, while those formed outside the clumps,
in the disk field where star formation proceeded at a slower rate, are $\alpha$\--poor. The model offers a qualitative interpretation of
the observed bimodality, although it is not sufficiently detailed to address the vertical segregation of the two components.

Finally, evidence of very m-poor stars (i.e., [Fe/H]$\leq-$2\,dex) in the inner Galaxy, with spatial and dynamical properties compatible 
with a pressure\--supported component, has been recently presented by \citet{PIGSVIII} by combining Gaia DR3 kinematics with the metallicity 
from the PIGS survey.

\section{Bulge GCs}\label{sec:GCs}

The GC system of the MW, as that of most giant galaxies, is bimodal in metallicity. The metal rich peak, at [Fe/H]$\sim$$-$0.8 dex, 
is known to be associated to the Galactic bulge, both spatially and dynamically \citep{barbuy+98,bica+16, perez-villegas+20}. 
Compared to their halo counterpart, bulge GCs are less studied, due to observational challenges such as large extinction and field 
contamination. For the same reason, the census of low-mass bulge GCs is not complete, with new, low-mass members being discovered 
almost yearly \citep[see][for a recent compilation]{bica+24}. Accurate study of their stellar content is important because GCs, as 
the closest incarnations of simple stellar populations, are the main calibrators of stellar evolution models, and only bulge GCs 
allow to extend these calibration to the m\--rich regime. 
Let us remember that, due to the well known mass-metallicity-age relation in galaxies, giant galaxies in the universe are dominated 
by old, solar and super-solar-metallicity stars. 

It should be noted that some of the GCs spatially and dynamically associated to the Galactic bulge are relatively m\--poor \citep[e.g.,][]{barbuy+2018}. They are especially interesting as they are expected to include the oldest GC in the Galaxy \citep{tumlinson+10}.

A couple of bulge GCs, namely Terzan~5 and Liller~1, deserve special mention as they show a large spread in age and metallicity, not 
seen before in any other GC \citep{ferraro+16_T5, ferraro+21_L1}. While their nature and origin is still not firmly established, the 
metallicity distribution and $\alpha$\--elements abundance of their sub\--populations are remarkably similar to what observed in bulge 
field stars.

\section{Summary}

With a stellar mass of M$_*\sim 1.7-2 \times 10^{10}$ M$_{\odot}$, mostly made of $\sim$10\,Gyr old stars, the MW bulge is the first 
massive component to form, and the only bulge that can be explored in great detail, hence its relevance in the context of galactic 
and extra\--galactic studies.

The bulge hosts a bar with an inclination of $\sim$27 degrees with respect to the Sun\--Galactic center direction,extending for about 
1\,kpc from the center. Its semi\--minor axis ratios are close to a:b/a:c/a=1:0.4:0.3.

The dynamical evolution of the bar (i.e.,  buckling and vertical resonance heating) lends the bulge a boxy/peanut shape in its outer 
regions.

Overall, the bulge stellar population is rich in iron, but its sub\--solar (i.e., m\--poor) and super\--solar (i.e., m\--rich) 
components display very distinct properties in terms of spatial distribution, kinematics and abundances pattern. The m\--rich 
component is depleted in $\alpha$\--elements and is spatially arranged in a bar\--like structure. On the other hand, the m\--poor 
and $\alpha$\--rich component does not appear to follow the bar, being symmetrically concentrated as a spheroid\--like structure.

Although the bulge as a whole rotates cylindrically, m\--poor stars show slower rotation with a relatively large velocity dispersion 
($\sigma_{\rm RV}$=100\--120 km/s) rather constant across the sky. The m\--rich component, instead, has a velocity dispersion gradient, 
with values raging from $\sigma_{\rm RV}$=60 km/s at b$=-8^\circ$ to $\sigma_{\rm RV}$=140 at b=$-1^\circ$.

Finally, while the [$\alpha$/Fe] distribution of the m\--poor and m\--rich components strongly suggests an association with the thick 
and thin disk, respectively, the origin of a clear gap in between is not yet completely understood.




\begin{ack}[Acknowledgments]

M. Z. acknowledges financial support by the National Agency for Research and Development (ANID), through grants: FONDECYT Regular No. 1230731;  
the Millenium Science Initiative, ICN12\_009 and AIM23-0001, awarded to the Millennium Institute of Astrophysics  (MAS); and the BASAL Center for Astrophysics and Associated Technologies (CATA) grant FB210003.

E. V. acknowledges the Excellence Cluster ORIGINS Funded by the Deutsche Forschungsgemeinschaft (DFG, German Research Foundation) under Germany’s Excellence Strategy – EXC-2094-390783311. 
\end{ack}


\bibliographystyle{Harvard}
\bibliography{bulgechap}

\end{document}